# Relative In-flight Response of IBEX-Lo to Interstellar Neutral Helium Atoms

P. Swaczyna[1,*], M. Bzowski[2], S. A. Fuselier[3,4], A. Galli[5], J. Heerikhuisen[6], M. A. Kubiak[2], D. J. McComas[1], E. Möbius[7], F. Rahmanifard[7], N. A. Schwadron[1,7]

[1]Department of Astrophysical Sciences, Princeton University, Princeton, NJ 08544, USA
[2]Space Research Centre PAS (CBK PAN), Bartycka 18a, 00-716 Warsaw, Poland
[3]Southwest Research Institute, San Antonio, TX 78228, USA
[4]University of Texas at San Antonio, San Antonio, TX 78249, USA
[5]Physics Institute, University of Bern, Bern, 3012, Switzerland
[6]Department of Mathematics and Statistics, University of Waikato, Hamilton, New Zealand
[7]Physics Department, Space Science Center, University of New Hampshire, Durham, NH 03824, USA

**Abstract**

The IBEX-Lo instrument on the Interstellar Boundary Explorer (IBEX) mission measures interstellar neutral (ISN) helium atoms. The detection of helium atoms is made through negative hydrogen (H⁻) ions sputtered by the helium atoms from the IBEX-Lo conversion surface. The energy spectrum of ions sputtered by ISN helium atoms is broad and overlaps the four lowest IBEX-Lo electrostatic analyzer (ESA) steps. Consequently, the energy response function for helium atoms does not correspond to the nominal energy step transmission. Moreover, laboratory calibration is incomplete because it is difficult to produce narrow-energy neutral atom beams that are expected for ISN helium atoms. Here, we analyze the ISN helium observations in ESA steps 1–4 to derive the relative in-flight response of IBEX-Lo to helium atoms. We compare the ratios of the observed count rates as a function of the mean ISN helium atom energy estimated using the Warsaw Test Particle Model (WTPM). The WTPM uses a global heliosphere model to calculate charge exchange gains and losses to estimate the secondary ISN helium population. We find that the modeled mean energies of ISN helium atoms, unlike their modeled fluxes, are not very sensitive to the very local interstellar medium parameters. The obtained relative responses supplement the laboratory calibration and enable more detailed quantitative studies of the ISN helium signal. A similar procedure that we applied to the IBEX-Lo observations may be used to complement laboratory calibration of the next-generation IMAP-Lo instrument on the Interstellar Mapping and Acceleration Probe (IMAP) mission.

## 1. Introduction

Interstellar neutral (ISN) atoms from the interstellar medium are the primary source of neutral atoms in the heliosphere (Fahr 1968; Wallis 1975). While many ISN atoms are ionized before reaching the inner heliosphere, they are the primary interstellar material available for direct sampling with space instrumentation at 1 au from the Sun. One of the main objectives of the Interstellar Boundary Explorer (IBEX) mission (McComas et al. 2009) is to diagnose the ISN atoms in the heliosphere (Möbius et al. 2009b). The low-energy sensor on this mission – IBEX-Lo (Fuselier et al. 2009) – was designed to measure neutral atom fluxes from ~10 eV to 2 keV with eight logarithmically-spaced energy steps, including multiple ISN species (Möbius et al. 2009a).

Each ISN species is comprised of two populations. The first one, called the primary population, originates from far beyond the heliosphere boundaries, where the presence of the heliosphere does not modify the very local interstellar medium (VLISM). The VLISM is defined here as the interstellar medium around the heliosphere, which is distinct from the Local Interstellar Cloud and G-Cloud (Swaczyna et al. 2022b). This region encompasses both the pristine VLISM far from the heliopause, which is  not modified by the presence of the heliosphere, and the outer heliosheath, where the interstellar plasma is deflected (Galli et al. 2022). The primary ISN atoms decouple from the plasma and may cross the heliopause, but they are modulated by collisions. Most importantly, charge exchange between ions and atoms creates a secondary population that inherits properties from the disturbed plasma (Baranov et al. 1991). The secondary population is slower and warmer than the primary population because it is created from slowed-down and heated plasma outside the heliopause.

---

[*] Corresponding author (swaczyna@princeton.edu)

The most abundant ISN species at 1 au is helium because it is less likely than hydrogen to be ionized in the heliosphere as the atoms travel from the VLISM to 1 au (Rucinski & Bzowski 1995; Sokół et al. 2019a). Therefore, counts from ISN helium atoms dominate the observations in the four lowest energy steps of the IBEX-Lo instrument. The energy steps are defined by the throughput of the electrostatic analyzer (ESA) controlled by the electric potentials of the ESA surfaces. The preset values of these potentials define eight energy steps to which we later refer to as ESA steps, e.g., ESA 1 denotes the lowest routinely used energy step on IBEX-Lo (Fuselier et al. 2009).

The ISN helium observations are used to determine the physical properties of the pristine VLISM, such as the flow velocity and temperature, because this population is abundant and not affected much by interactions in the heliosphere. The IBEX-Lo observations of the primary ISN helium population were analyzed in a series of papers (Bzowski et al. 2012, 2015; Möbius et al. 2012, 2015; Leonard et al. 2015; Schwadron et al. 2015, 2022; Swaczyna et al. 2018, 2022a). In addition, the secondary population of ISN helium, initially dubbed the Warm Breeze, was also discovered, and for the first time measured, using IBEX observations (Kubiak et al. 2014, 2016; Park et al. 2016; Bzowski et al. 2017, 2019).

Quantitative analyses of the ISN helium observations use least-squares comparison between modeled ISN fluxes integrated with the instrument response function (Schwadron et al. 2015; Sokół et al. 2015b; Swaczyna et al. 2015) or determine moments of the observed signal and compare those with analytic predictions (Lee et al. 2012, 2015; Möbius et al. 2015). Initially, studies of ISN helium focused on analyses of observations from ESA 2 because it was noticed that, in this step, the signal from ISN hydrogen is significantly reduced compared to ESA 1 (Saul et al. 2012; Galli et al. 2019). Moreover, it was assumed that the energy response in ESA 2 can be approximated as constant over the energy of atoms coming into the instrument. In later studies, Swaczyna et al. (2018, 2022a) included ESA 1 and 3 in the analysis of ISN helium observations, assuming a linearly changing response function in all three energy steps. Nevertheless, the linear function within each step is insufficient to describe the energy response of the instrument fully. Figure 1 presents IBEX-Lo observations of ISN helium in ESA 1–4 from observation seasons 2013–2020 projected in the sky (Swaczyna et al. 2022a). The data range is the same as that used by Bzowski et al. (2019). The figure shows that the signal near the peak of ISN helium increases gradually from ESA 1 to 3 and drops significantly in ESA 4. Moreover, while the peak in ESA 3 is higher than in ESA 2, it is evident that the portion dominated by the Warm Breeze in the upper right part of the data shown is reduced.

In this paper, we present a comprehensive analysis of the ISN helium observations from IBEX-Lo in ESA 1–4 to find the relative calibration of this instrument capable of explaining the observed differences between the ESA steps. This instrument response is needed to improve the accuracy of quantitative studies of ISN helium observations from IBEX, especially if they use a broader data range covering both ISN populations, which results in a broader range of helium atoms in the IBEX reference frame. Section 2 justifies using the relative response function for ISN helium atoms. Later, we discuss the ISN helium atom energy estimation in the IBEX frame (Section 3) and the fitting of the response function using IBEX observations (Section 4). The response function obtained from this procedure is presented in Section 5, and the robustness of the derivation is discussed in Section 6. Finally, we summarize our results and present perspective for future instruments in Section 7.

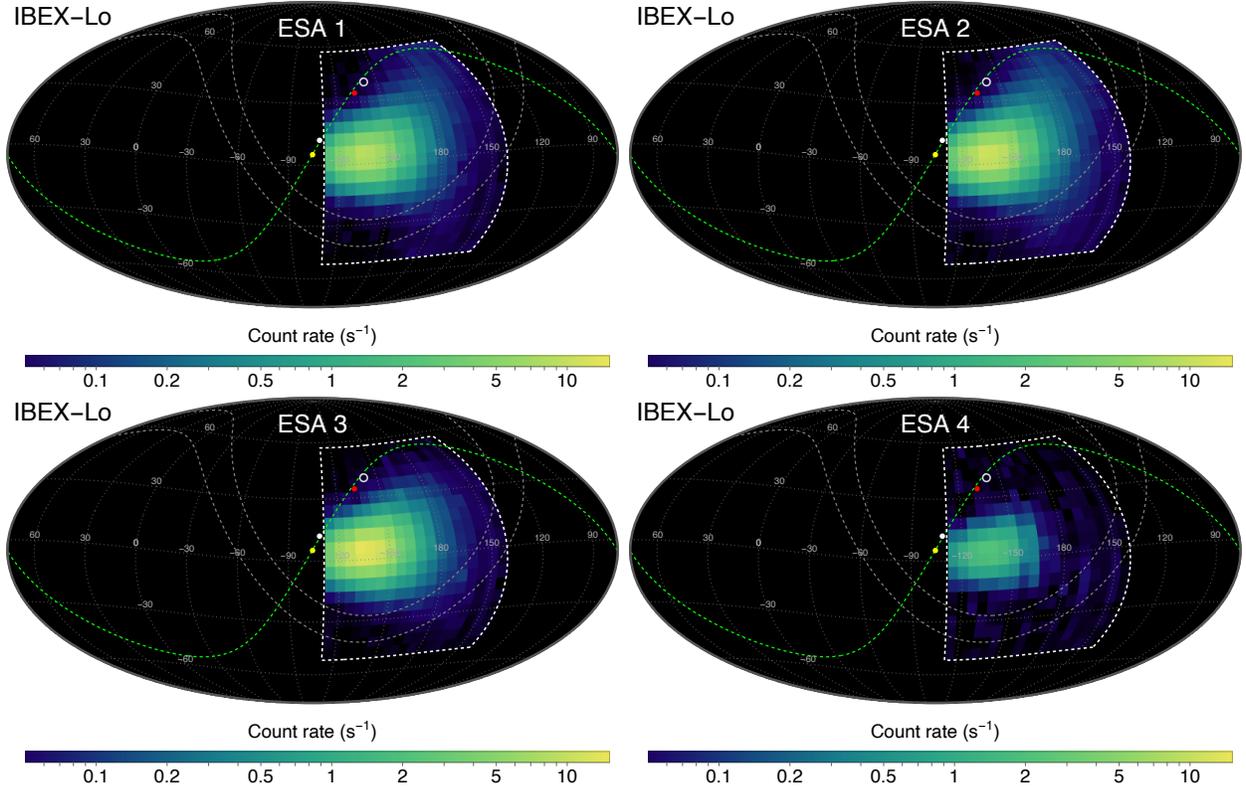

**Figure 1.** Skymaps of IBEX-Lo observations of ISN helium atoms collected in ISN seasons 2013-2020 in ESA 1–4 (top left to bottom right panels). The presented data are limited to the range dominated by ISN helium atoms (dashed white polygon). The green dashed line represents the neutral deflection plane, including the pristine VLISM flow (yellow dot), Warm Breeze flow (white dot), pristine VLISM magnetic field (red dot), and IBEX ribbon center (gray circle) directions. The outline of the IBEX ribbon is marked with dashed gray lines.

## 2. IBEX-Lo Relative Response for ISN Helium Atoms

The direct mechanism for neutral atom detection in IBEX-Lo depends on the negative ionization of atoms scattered off from conversion surfaces (Scheer et al. 2006; Wurz et al. 2006, 2008; Wieser et al. 2007). IBEX-Lo uses a diamond-like carbon conversion surface oriented at a shallow 15° angle relative to the incoming atom trajectories (Fuselier et al. 2009). Some of the incoming atoms are ionized and reflected off the conversion surface into the ESA, which selects the energy range of negative ions that can reach the detector section. In the detection section, the ions are accelerated with post-acceleration (PAC) voltage and measured using a coincidence time-of-flight (TOF) section. The energy is determined based on the ESA setting, while the TOF section allows for measuring the mass. The energy of ions reflected off the conversion surface is reduced due to scattering by about 20%–30% compared to impacting atoms. This loss is accounted for in the definition of the energy steps, which have central energies of incident H atoms of 15, 29, 55, and 110 eV for ESA steps from 1 to 4 with an energy passband of $\Delta E/E \approx 0.7$ (Möbius et al. 2009a; Schwadron et al. 2022)

ISN hydrogen atoms are subject to radiation pressure, and thus they are not significantly accelerated inside the heliosphere. Consequently, their typical speed relative to the instrument peaks at ~55 km s$^{-1}$, a sum of their speed relative to the Sun (~25 km s$^{-1}$) and the Earth's orbital motion of ~30 km s$^{-1}$. This speed corresponds to an energy of ~16 eV. Therefore, ISN hydrogen atoms are observed in ESA 1 and 2. ISN helium atoms are accelerated in the heliosphere by the Sun's gravity, and thus their speed in the spacecraft frame can reach up to ~80 km s$^{-1}$, corresponding to an impact energy of ~134 eV. Consequently, one could expect the dominant ISN helium signal in ESA 4 and 5. However, because these atoms do not produce stable negative ions, they are instead observed through their sputtered products, mainly in ESA 1–3, while the measured event rates in ESA 4 are an order of magnitude smaller (Section 3). Because of the sputtering, the observed events are observed to be H$^-$ ions in the TOF section. The observations in the lower ESA steps than the incoming atom energy are caused by the broad energy spectrum of ions sputtered by helium atoms off the conversion surface. Like everything else on the spacecraft, the conversion surface is contaminated with a residual, largely water layer on top of the diamond-like carbon surface. Consequently, helium atoms sputter negative H$^-$ and O$^-$ ions in addition to C$^-$ ions from the conversion surface, which are then detected in these energy steps. At the same time, the yield of negative He$^-$ ions is very low, and those ions are only metastable with lifetimes too short to reach the detector section (Wurz et al. 2008). Figure 2, reproduced from Schwadron et al. (2022), shows a schematic distribution of the sputtered and converted ions relative to the ESA passbands. While the distribution of converted ions is only slightly shifted in energy, the sputtered ions form a spectrum across all lower ESA steps. Because the energy distribution of sputtered ions crosses over multiple ESA steps, even for a narrow distribution of incoming ISN helium atoms, these ions are observed in all ESA from 1 through 4.

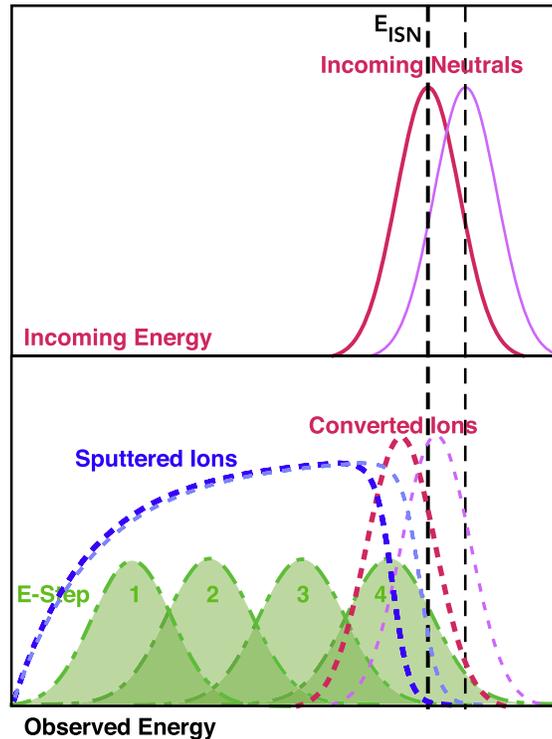

**Figure 2.** Schematic view of energy distributions of incoming atoms (top panel) and observed sputtered and converted ions compared with IBEX-Lo ESA passbands (bottom panel). Figure reproduced under the Creative Commons Attribution License CC BY 4.0 from Schwadron et al. (2022).

To derive the relative response function of ESA steps 1–4 in the IBEX-Lo instrument, we need to consider how the subsystems contribute to the energy response of the instrument. The entrance subsystem does not impact the energy response because it only filters out charged particles and limits the field of view using a physical collimator. Inside the instrument, neutral helium atoms hit the conversion surface. The sputter yield of negative hydrogen ions emitted from the surface is denoted here as $Y(E_{\text{atom}})$, where $E_{\text{atom}}$ is the energy of the impacting atom. Galli et al. (2015) simulated this yield using the SRIM simulation package (Stopping and Range of Ions in Matter, Ziegler et al. 2010). They found that this yield is at most ~0.5 for the considered energy range.

Because the ESA selects the energy of sputtered ions, we need the probability distribution of sputtered ions as a function of energy. In the most general form, this probability distribution is a function of $E_{\text{atom}}$ and of the sputtered ion energy $E_{\text{ion}}$: $S(E_{\text{atom}}, E_{\text{ion}})$. Here, we will make the first important assumption in our derivation that this function can be represented as a function of the ratio ($\eta = E_{\text{ion}}/E_{\text{atom}}$) of the ion to atom energies:

$$S(E_{\text{atom}}, E_{\text{ion}}) = s\left(\frac{E_{\text{ion}}}{E_{\text{atom}}}\right) = s(\eta). \tag{1}$$

This assumption may be validated through laboratory experiment or simulations. Because the function $s(\eta)$ represents a probability distribution, it is normalized as: $\int_0^\infty s(\eta)d\eta = 1$. While the function should be zero for $\eta > 1$ because the sputtered ions must have lower energies than the impacting atoms, this property is not needed for the derivation presented in this section.

Sputtered ions entering the ESA have a probability $T_{\text{ESA}e}(E_{\text{ion}})$ to be transmitted to the TOF section of IBEX-Lo in ESA step $e$. This transmission function scales with the central energy of the ESA steps ($E_{\text{ESA}e}$) and thus can be represented using a universal function $t(x)$ for all ESA steps in the following way:

$$T_{\text{ESA}e}(E_{\text{ion}}) = t\left(\frac{E_{\text{ion}}}{E_{\text{ESA}e}}\right) = t\left(\eta \frac{E_{\text{atom}}}{E_{\text{ESA}e}}\right), \tag{2}$$

where the second equality uses the definition of the ion-to-atom energy ratio $\eta$. Here, we can also assume that function $t(x)$ is formally defined from 0 to $\infty$ even though it is non-zero over a limited range of energy ratios.

The final element of the helium atom measurement in IBEX-Lo is the detection in the TOF section. After the ions exit the ESA, they are accelerated by the PAC voltage to reduce the energy loss and scattering of the ions and to increase the yields of electrons emitted from two foils through which the ions strike, located in the detector section. Thus, the efficiency of ion detection is increased. The second important assumption that we make in this derivation is that the detection efficiency in the TOF section $H_p(E_{\text{ion}}/E_{\text{atom}}, E_{\text{atom}}/E_{\text{ESA}e}, E_{\text{ion}}/E_{\text{ESA}e})$ depends on the PAC voltage $p$ and the energy ratios: $E_{\text{ion}}/E_{\text{atom}}$, $E_{\text{atom}}/E_{\text{ESA}e}$, and $E_{\text{ion}}/E_{\text{ESA}e}$, but not explicitly on the energy of ions or atoms. For the energy range of incoming ISN helium atoms, the atom energy in the TOF section is determined by the PAC voltage rather than by the impacting atom energy. This justifies the assumption that the TOF efficiency does not directly depend on the atom energy at the instrument entrance. Because only two of these ratios are independent, this can be expressed in a more straightforward form:

$$H_p\left(\frac{E_{\text{ion}}}{E_{\text{atom}}}, \frac{E_{\text{atom}}}{E_{\text{ESA}e}}, \frac{E_{\text{ion}}}{E_{\text{ESA}e}}\right) = h_p\left(\eta, \frac{E_{\text{atom}}}{E_{\text{ESA}e}}\right). \tag{3}$$

The dependence of the detection efficiency on the energy ratios describes the collimation property of the post-acceleration voltage. Ions sputtered off the conversion surface have some angular distribution of directions when escaping from the surface. Similarly, ions transmitted through the ESA have trajectories escaping at different angles depending on the ratio of the ion energy to the ESA central energy. The ions with energies close to the central energy exit ESA at similar angle to the one at which they entered the ESA.

Based on the above components of the response function, the probability $q_{p,e}(E_{\text{atom}})$ that an atom is detected in ESA step $e$ under PAC voltage $p$ is given by the integral over possible ion-to-atom energy ratios $\eta$:

$$q_{p,e}(E_{\text{atom}}) = \int_0^\infty Y(E_{\text{atom}}) s(\eta) t\left(\eta \frac{E_{\text{atom}}}{E_{\text{ESA}e}}\right) h_p\left(\eta, \frac{E_{\text{atom}}}{E_{\text{ESA}e}}\right) d\eta$$

$$= Y(E_{\text{atom}}) \underbrace{\int_0^\infty s(\eta) t\left(\eta \frac{E_{\text{atom}}}{E_{\text{ESA}e}}\right) h_p\left(\eta, \frac{E_{\text{atom}}}{E_{\text{ESA}e}}\right) d\eta}_{\equiv f_p\left(\frac{E_{\text{ESA}e}}{E_{\text{atom}}}\right)} = Y(E_{\text{atom}}) f_p\left(\frac{E_{\text{ESA}e}}{E_{\text{atom}}}\right). \quad (4)$$

The yield $Y(E_{\text{atom}})$ of sputtered particles is taken outside the integral, as it does not depend on the ratio $\eta$. The remaining part of the integral depends only on the ratios $\eta$ and $E_{\text{atom}}/E_{\text{ESA}e}$. While we do not need to know the specific form of the functions in the integrand, the dependency on $\eta$ is integrated over. Thus we can define a new function $f_p(E_{\text{ESA}e}/E_{\text{atom}})$ depending only on the other ratio. We further refer to this function as the relative response function, and we aim to find it in this paper.

### 3. Energy of ISN Helium Atoms in the IBEX Frame

We need to estimate the mean speed of ISN helium atoms entering the IBEX-Lo instrument to find the relative response function. As ISN helium atoms are transported from the VLISM to 1 au, they gain speed in the solar gravitational field, and radiation pressure is negligible. The conservation of energy shows that the atom's speed at distance $r$ from the Sun is

$$v(r) = \sqrt{\frac{2GM_\odot}{r} + v_\infty^2}, \quad (5)$$

where $G$ is the gravitational constant, $M_\odot$ is the solar mass, and $v_\infty$ is the atom's speed at infinity. For the typical ISN helium atom speed at infinity of ~25 km s⁻¹, this formula gives the speed of ~49 km s⁻¹ at $r = 1$ au. Moreover, for this typical speed, the thermal spread of speeds reduces by about a factor of two: $dv(1\text{ au}) \approx 0.5 dv_\infty$.

The second element of the atom speed relative to IBEX is the spacecraft velocity around the Sun. Because the spacecraft is in an orbit around the Earth, the main component of this speed is the Earth's orbital speed of ~30 km s⁻¹. The IBEX-Lo boresight tracks a great circle perpendicular to the spin axis, which is repointed every few days to approximately follow the Sun. Considering idealized conditions in which the spacecraft moves on a circular orbit around the Sun at the distance of 1 au, i.e., neglecting the spacecraft motion around the Earth, the highest relative velocity for atoms coming from a different direction at the same speed in the solar frame occurs when the instrument boresight is in the ecliptic plane and thus is aligned with the spacecraft velocity. As the boresight moves away from the ecliptic, the relative velocity decreases because the velocity vectors are not aligned.

The periodic repointing of the spacecraft changes the boresight's ecliptic longitude. For observations at different ecliptic longitudes, the mean atom speed at IBEX depends on the portion of the distribution function observed for those directions. The peak position longitude $\lambda_{\text{Peak}}$ due to gravitational deflection of ISN helium atoms is shifted from the inflow longitude $\lambda_{\text{Inflow}}$ according to the following formula (Lee et al. 2012; Möbius et al. 2012):

$$\lambda_{\text{Inflow}} - \lambda_{\text{Peak}} = \arcsin \frac{GM_\odot}{GM_\odot + r_0 v_\infty^2} \approx 35°, \quad (6)$$

where $r_0 = 1$ au. While the peak of the distribution is shifted, faster ISN helium atoms are observed closer to the inflow direction because they are less deflected. While the flux decreases when pointing closer to the pristine VLISM inflow direction, we observe a part of the distribution that is faster, and therefore, the observed mean speed is higher. In the presented analysis, the model fluxes are not needed as we only use relative count rates between ESA step pairs.

To find the quantitative shapes of the atoms' distributions under the above assumptions, we solve the equation for Kepler trajectories of atoms of different speeds at IBEX under the above-described idealized conditions for several directions in the ecliptic plane and a few directions out of the ecliptic at the ecliptic longitude where the peak of the incoming ISN flow is observed. Figure 3 presents the distribution function of atoms in the VLISM as a function of their speed in the spacecraft frame. The figure confirms that the maximum mean speed is expected for atoms coming from the proximity of the ecliptic plane. For different points in the ecliptic plane, the highest mean speed is expected for atoms with velocities closer to the inflow direction.

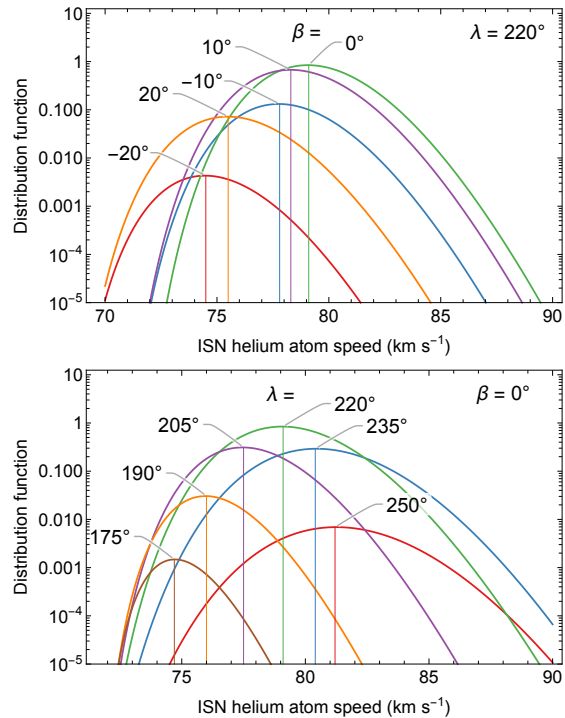

**Figure 3.** Modeled distribution function of ISN helium atoms along different directions in the sky of the IBEX-Lo boresight under idealized conditions assuming Sun-pointing of IBEX (cf. text). Top panel presents results for ecliptic longitude $\lambda = 220° \approx \lambda_{\text{Peak}}$ and latitudes $\beta = -20°, -10°, 0°, 10°, 20°$. The bottom panel presents results in the ecliptic plane ($\beta = 0°$) and longitudes $\lambda = 175°, 190°, 205°, 220°, 235°, 250°$.

The above discussion accounts for only the primary population of ISN helium atoms under idealized observation conditions. However, to precisely estimate the energy of incoming atoms, we need to account for the spacecraft state vector and how it changes during each observed orbit/arc. Therefore, we use the Warsaw Test Particle Model (WTPM, Sokół et al. 2015b; Bzowski et al. 2017, 2019) to integrate the mean speed of atoms observed at each IBEX-Lo data point. We use data points averaged over the good time intervals for each orbit/arc number and spin angle bin. This model calculates the mean speed and the mean squared speed of the ISN atoms as moments of the ISN helium atom distribution at IBEX.

In this study, we calculate the combined ISN helium populations following the methodology presented by Bzowski et al. (2017, 2019), which uses a global heliosphere model (Zirnstein et al. 2016; Heerikhuisen et al. 2019) to calculate gains and losses to the ISN helium population due to charge exchange between helium atoms and ions in the outer heliosheath producing the secondary population and is also causing losses to the primary population. We use the same pristine VLISM and solar wind conditions as in Swaczyna et al. (2023). The WTPM accounts for charge exchange collisions but neglects angular scattering in those collisions and elastic collisions, which result in the slowdown and heating of the ISN helium atoms in the outer heliosheath (Swaczyna et al. 2021). We will discuss this effect further in Section 6.

Figure 4 shows the flux, mean speed, and standard deviation of the speed obtained from the WTPM and projected onto the sky following the same procedure that was used to project the IBEX data in Figure 1. Consistent with the above discussion, the highest mean speed is observed close to the ecliptic plane for larger ecliptic longitudes (close to the left edge of the ISN helium data range shown in those plots). The speed could be even higher for longitudes beyond the data range shown here, but the portion at larger longitudes includes a significant contribution from ISN hydrogen atoms (Galli et al. 2019) and thus is not suitable for our analysis, which assumes a single source of the observed signal.

We also find that the speed distribution of ISN helium atoms is narrow, represented by a standard deviation of less than ~4 km s$^{-1}$ and typically within the range of 2–3 km s$^{-1}$. This effective width of the distribution includes contributions from (1) the thermal spread of the ISN helium distribution function at 1 au, (2) broadening due to the spacecraft velocity change over the good time intervals, and (3) integration over the part of the sky included in each data bin. Nevertheless, the distribution of ISN helium is narrow enough that it can be considered mono-energetic for our analysis.

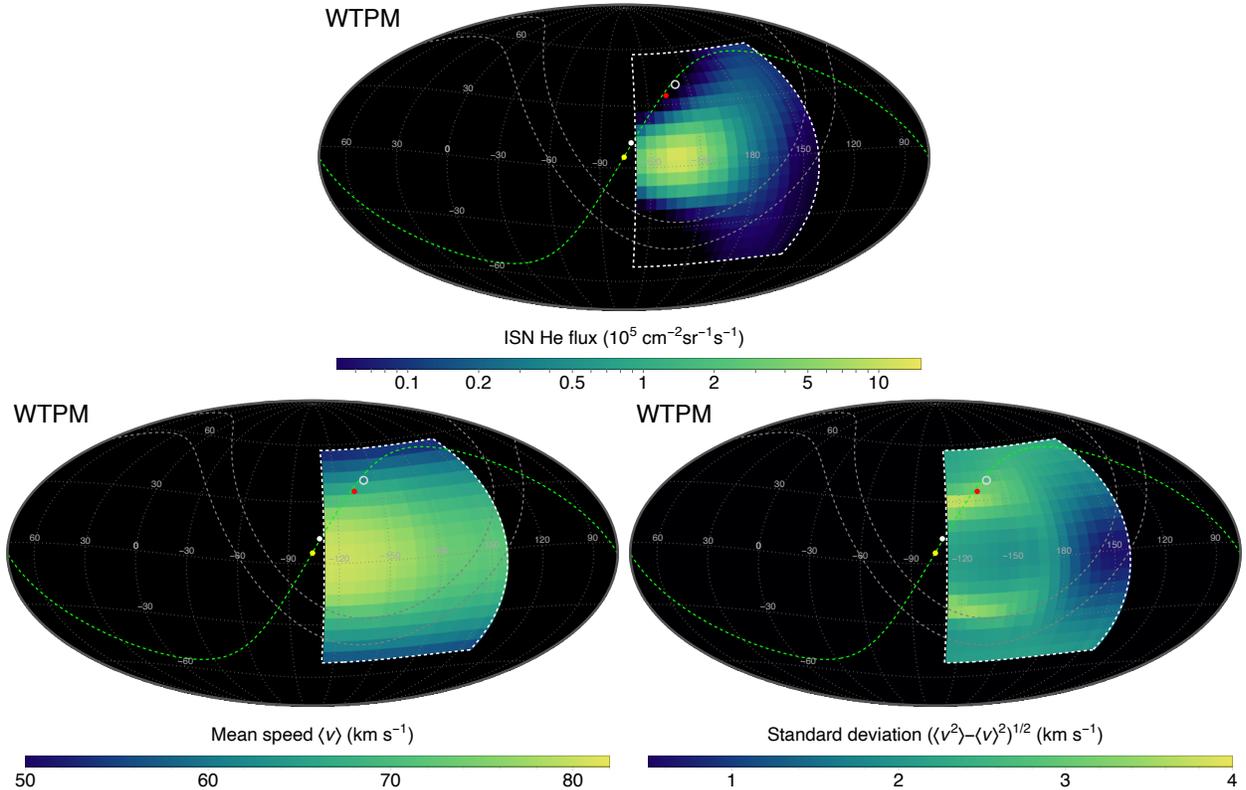

**Figure 4.** Modeled ISN helium flux (top panel), mean speed (bottom left panel), and standard deviation of the modeled distribution (bottom right panel) from the WTPM for the ISN helium observation season projected on the sky. The model accounts for gains and losses due to charge exchange in the outer heliosheath.

## 4. Fitting Procedure for the Relative Response of IBEX-Lo ESA steps

ISN helium count rates observed by IBEX-Lo over a range of angle bins in ESA step $e$ are an integral of the energy response function given in Equation (4) with the differential flux of ISN helium atoms $j(E)$, which under the assumption of a mono-energetic beam is:

$$c_{p,e} = G \int_0^\infty q_{p,e}(E) j(E) dE \approx G q_{p,e}(E_{\text{atom}}) J = G Y(E_{\text{atom}}) f_p\left(\frac{E_{\text{ESA}e}}{E_{\text{atom}}}\right) J \quad (7)$$

where $J = \int_0^\infty j(E)\, dE$ is the integral flux of atoms incoming to the instrument, $G$ is the effective geometric factor accounting for the opening of the instrument collimator, but that does not include the energetic response included in the response function, and $E_{\text{atom}}$ is the energy of atoms observed in each bin. Consequently, the ratio of the count rates observed in two ESA steps $j$ and $k$ is equal to the ratio of the relative response function $f_p(x)$:

$$r_{j,k} \equiv \frac{c_j}{c_k} = \frac{Y(E_{\text{atom}}) f_p(E_{\text{ESA}j}/E_{\text{atom}}) J}{Y(E_{\text{atom}}) f_p(E_{\text{ESA}k}/E_{\text{atom}}) J} = \frac{f_p(E_{\text{ESA}j}/E_{\text{atom}})}{f_p(E_{\text{ESA}k}/E_{\text{atom}})}. \quad (8)$$

This equation provides the means to find the relative response function using ratios of observed count rates in different ESA steps together under the assumption that the energy of observed atoms is obtained from the model. No explicit knowledge of the atom flux into the instrument is needed for this method.

The form of the response function is not known a priori, however, we expect it to be a relatively smooth function because the various internal processes in the measurement vary smoothly with particle energy. Here we assume that the response function can be approximated by a polynomial in log-log space:

$$f(x) = A \exp\left(\sum_{i=1}^K m_i (\ln x)^i\right) \quad (9)$$

where $m_i$ are coefficients of the sought polynomial, $A$ is the normalization factor, and $K$ is the degree of the polynomial. For brevity, we skip subscript $p$ denoting possible PAC voltage settings, but we perform separate analyses for the two PAC voltages used during the IBEX-Lo mission. We omit a constant in the sum because it is equivalent to a change of the normalization factor. With the above definition, we find the polynomial coefficients by minimizing the following least-squares sum:

$$\chi^2(m_1, m_2, \ldots, m_n) = \sum_{l \in \text{bins}} \sum_{(j,k) \in \text{ESA pairs}} \frac{\left(r_{l,j,k} - \frac{f(E_{\text{ESA}j}/E_{\text{atom},l})}{f(E_{\text{ESA}k}/E_{\text{atom},l})}\right)^2}{\delta r_{l,j,k}^2}, \quad (10)$$

where we use subscript $l$ to enumerate different data bins enumerated by orbit or arc number and spin angle bin. The second sum is over all possible pairs of the four lowest energy steps. Only three different ratios are statistically independent, e.g., $(j,k) \in \{(1,2), (1,3), (1,4)\}$, while the other pairs are quotients of these. In our analysis, we do account separately for all possible pairs, including the reversed order of ESA steps, e.g., we include pair $(j,k) = (1,2)$ as well as $(j,k) = (2,1)$. Consequently, we have 12 possible pairs of ESA steps included in the sum. While this procedure leads to double counting, it minimizes the impact of non-Gaussian uncertainties of the ratios deduced from the Poisson uncertainties in the count rates.

The ratio of the count rates in bin $l$ between ESA step $j$ and $k$ is thus denoted $r_{l,j,k}$ with its uncertainty $\delta r_{l,j,k}$. For this analysis, we calculate the rate in each bin using the following formula:

$$c_{l,e} = \frac{\gamma_{l,e} d_{l,e}}{t_{l,e}} - b_e, \quad (11)$$

where $\gamma_{l,e}$ is the throughput correction factor (Swaczyna et al. 2015, 2022a), $d_{l,e}$ is the number of registered counts, $t_{l,e}$ is the exposure time, and $b_e$ is the ubiquitous background rate (Galli et al. 2015, 2017). To calculate the uncertainty of the count rate, we account for the Poisson uncertainty: $\delta c_{l,e} = \gamma_{l,e}\sqrt{d_{l,e}}/t_{l,e}$. We neglect the uncertainties of the throughput corrections in this analysis because the main component of their uncertainties is highly correlated between ESA steps due to the common source of background events in these ESA steps, and thus the uncertainties reduce when we calculate the ratio of the count rates. We also neglect the uncertainties of the background rate because their variabilities are negligible compared to the Poisson uncertainties in the used data range. The uncertainties of the count rate ratios are the propagated uncertainties of the rates:

$$\delta r_{l,j,k} = \left[\left(\frac{\delta c_{l,j}}{c_{l,j}}\right)^2 + \left(\frac{\delta c_{l,k}}{c_{l,k}}\right)^2\right]^{1/2} r_{l,j,k}. \quad (12)$$

The $\chi^2$ minimization is performed separately for the two PAC voltages used during the IBEX mission. The nominal PAC voltage (16 kV) was used until the end of the ISN season in 2012. Then, between ISN seasons 2012 and 2013, the PAC voltage was reduced to 7 kV, which reduced the detection efficiency approximately by half. Moreover, between these two seasons, the TOF logic was changed, which eliminated the buffer losses, and thus the throughput correction is not needed for later observation seasons.

Our analysis uses observation periods that covered at least the three lowest ESA steps. This criterion eliminates data collected in the Oxygen and cross-calibration modes during which ESA 3 is not used (see Swaczyna et al. 2022a). Consequently, the data collection under the 16 kV PAC voltage is limited to seasons 2009 and 2010 and two arcs in 2012: 156b and 157a. The data set collected under 7 kV PAC voltage includes ISN seasons 2013-2020, except for season 2016, during which ESA 3 was not observed. While we use ESA 4 observations whenever available, we do not require these observations to include the ratios between ESA 1–3. To eliminate low count rate points, we use only data points where the mean count rate in ESA 1–3 is at least 0.1 s$^{-1}$, i.e., where they are by at least an order of magnitude higher than the background rates. This criterion minimizes the impact of unknown backgrounds. For ESA 4, we use a different criterion. Namely, we only use ratios with the ESA 4 rates if $E_{ESA4}/E_{atom,l} > 0.95$. Above this ratio, the rates are comparable with the background even if the mean rate in ESA 1–3 is above 0.1 s$^{-1}$. This criterion eliminates very low count rates possible in this ESA step because the signal is significantly reduced.

In our first fitting attempt, we use the data selection as in Bzowski et al. (2019). The selection is defined based on ecliptic longitudes with spin axis pointing from 235° to 335°, which corresponds to ecliptic longitudes of the boresight in the ram direction from ~145° to ~245°. This range includes the ISN helium peak at $\lambda_{Peak} \approx 220°$, see Equation (6). In addition, we use spin angle bins from 216° to 318°, which corresponds to ecliptic latitudes from –48° to +54°. This data range corresponds to the data shown in Figure 1. However, we found that the reduced $\chi^2$ value obtained in the minimization of Equation (4) of 1.67 for the PAC voltage of 16 kV. Because the part of the data close to the inflow direction includes a contribution from ISN hydrogen atoms (Swaczyna et al. 2018; Galli et al. 2019), we decided to use a reduced longitude range of the spin axis from 235° to 325°. As a result, the reduced $\chi^2$ for the optimal polynomial, with $n = 6$ coefficients, decreases that value to 1.13 and from 1.17 to 1.08 for the PAC voltage of 7 kV. The more significant reduction under the higher voltage is not surprising because the ISN hydrogen signal was very strong in seasons 2009 and 2010, which are the dominant ISN helium data collected with the PAC voltage of 16 kV. Due to special modes applied in 2011 and 2012, only two arcs in 2012 are included in the dataset as described in the previous paragraph.

To find the degree of the polynomial needed to model the response function, we performed the minimization for the maximum degree ranging from $K = 1$ to 10. Figure 5 presents the reduced $\chi^2$ as a function of $K$. The best-fit polynomials for $K = 1$ or 2 represent poor fits with $\chi^2 > 15N$, where $N$ is the number of data points. For higher-degree polynomials, the goodness of fit greatly improves. The higher the degree of the polynomial, the lower $\chi^2$. For the PAC voltage of 7 kV ($N = 7063$), which we will use to justify the optimal degree of the polynomial, the $\chi^2$, as shown in the figure, decreases significantly until $K = 5$ with only slight improvements for higher degrees. The polynomial is only an approximation of the true response function. Therefore, one may suggest using the Bayesian Information Criterion (BIC, Schwarz 1978; Liddle 2007) defined as BIC $= \chi^2 + K \ln N$ to select the optimal model. This criterion suggests that a model with $K = 8$ is optimal. However, we noticed that for this polynomial $m_8 > 0$, i.e., the model predicts that the response function increases both for high and low energy ratios $x \gg 1$ and $x \ll 1$, while we would generally expect it to drop to 0. To the contrary, the best-fit polynomial for $K = 6$ has $m_6 < 0$, i.e., the response function approaches 0 for high and low energies. Additionally, all polynomials with odd degrees result in unlimited values for the response at either of those limits. The relative change of the response function model for the polynomials with $K = 6$ and $K = 8$ over the range of the data does not exceed 3%, which is comparable with the uncertainty of this function related to the calculation of the incoming atom energy (see Section 6). Therefore, we use $K = 6$ polynomials to model the response function.

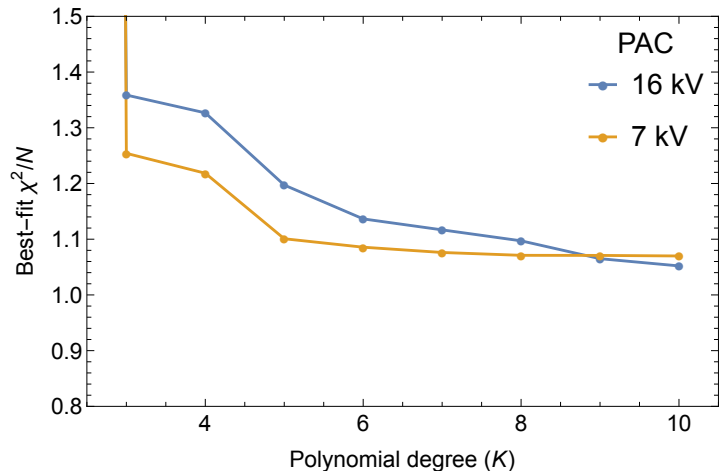

**Figure 5.** Best-fit $\chi^2/N$ as a function of the maximum polynomial degree $K$ for the two PAC voltages used during the mission. The datasets used here do not include the longitude range that likely includes a contribution from ISN hydrogen.

## 5. Best-fit Relative Calibration for ISN Helium Atoms

The fitting procedure for the response function presented in Section 4 is based on the ratios of this function in two different ESA steps for various energies of the incoming atoms in the spacecraft frame. Consequently, the normalization factor $A$ given in the response function provided in Equation (9) cannot be obtained from the fitting. Instead, we constrain the normalization factor so that the maximum of the response function is 1. The absolute calibration must be obtained from the laboratory calibration to supplement this relative response function. Table 1 shows the best-fit polynomial coefficients and the normalization factor defined above. The precision of numbers provided in the table allows for the reproduction of the best-fit response functions better than 0.03% over the range of the data.

**Table 1. Parameters of the relative response function given in Equation (9)**

| PAC | $m_1$ | $m_2$ | $m_3$ | $m_4$ | $m_5$ | $m_6$ | $A$ |
|---|---|---|---|---|---|---|---|
| 16 kV | −6.79873 | −8.41905 | −9.63405 | −8.22097 | −3.53226 | −0.564158 | 0.062703 |
| 7 kV | −5.82750 | −5.24562 | −4.99331 | −4.89035 | −2.36145 | −0.402573 | 0.070905 |

The numbers of the different ratios that we use for the fitting at PAC voltages of 7 kV and 16 kV are 7063 and 1957, respectively. While all these points are used individually for the fitting expression given in Equation (10), we decided to bin them based on the incoming atom energy into 1-eV-wide bins to make plots showing the fitting results. Based on the selection criteria stipulated in the previous section, this energy is within the range of 60 eV to 135 eV. However, some energy bins do not have any data points, especially for lower energies, because the mean energy within bins at high latitudes have similar energies regardless of the longitude, while bins in neighboring latitude bands have substantially different mean velocities (see the bottom left panel in Figure 2).

Figure 6 shows the ratios of the best-fit response function for different pairs of ESA steps and the binned observed ratios as a function of the incoming atom energy. Each curve representing the model response ratio is obtained from the same function. Therefore, they are not independent fits to each ratio. Because the rate ratio of ESA $k$ to ESA $j$ is an inverse of the ratio of ESA $j$ to ESA $k$, we only present ratios of higher to lower ESA steps to lower in this figure. The figure shows that the data uncertainties, calculated from Equation (12), are the smallest for higher energies because the bins with the highest count rates are limited to the ecliptic plane and thus have higher energies, as discussed in Section 3. The consistency of the observed rates with the best-fit curves confirms the consistency indicated by $\chi^2$ values close to the number of degrees of freedom.

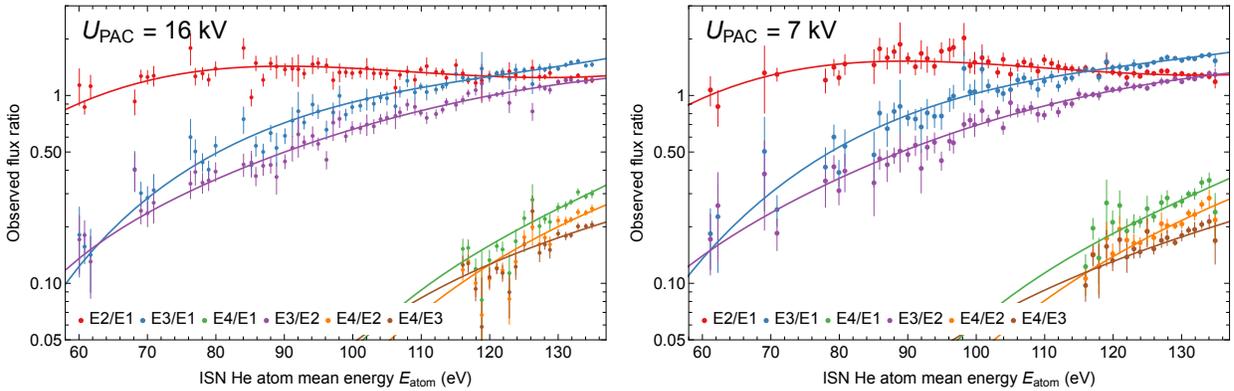

**Figure 6.** Ratios of the best-fit relative response function obtained from the model compared with the observed rate ratios as a function of incoming ISN helium atom mean energy. The observed rate ratios are binned in 1-eV-wide energy bins. The left and right panels show results for a PAC voltage of 16 and 7 kV, respectively.

While the response function can be characterized from the observed count rate ratios, we also want to compare this response with observations in each ESA step. This comparison requires an estimation of the ISN helium flux. For this purpose Schwadron et al. (2022) used ESA 3, assuming that the response function is constant for ISN helium atoms observed in this step. However, we found that this function significantly varies over this energy step. Therefore, we use a different approach using ESA steps 1–3 to estimate this flux. The relative response function allows us to combine different ESA steps to find the product of the geometric factor, the yield of sputtered particles, and the flux of incoming ISN helium atoms using an average based on Equation (7):

$$\langle GY(E_{\text{atom}})J\rangle = \frac{1}{3}\sum_{e=1}^{3}\frac{c_{p,e}}{f_p(E_{\text{ESA}e}/E_{\text{atom}})}. \tag{13}$$

We use this average to estimate the response function from observations in each ESA step as follows:

$$\tilde{f}_p(E_{\text{ESA}e}/E_{\text{atom}}) = \frac{c_{p,e}}{\langle GY(E_{\text{atom}})J\rangle}. \tag{14}$$

The averaged product in the denominator in Equation (14) is given by Equation (13), which is obtained directly from the observations. Therefore, we do not need to use the simulated flux for this calculation. The relative response function obtained from the fitting procedure is compared with these estimates in Figure 7. To calculate the product given in Equation (13), we only use the rates in ESA 1 to 3 because ESA 4 rates are reduced by order of magnitude. Therefore, they have a much lower statistical significance. The uncertainties in this figure are calculated from the uncertainty of the count rate calculated as discussed in Section 4. It is important to note that this figure could not be used to fit the model because it requires knowledge of the response function. Using vertical lines in these plots, we showed the minimum ratio (i.e., corresponding to the highest mean energy of ISN helium atoms) of energies for individual ESA steps, where the data points have the lowest uncertainties. The figure shows small overlaps between different ESA steps. Unfortunately, the highest ratio data points for each ESA step typically have much higher uncertainties.

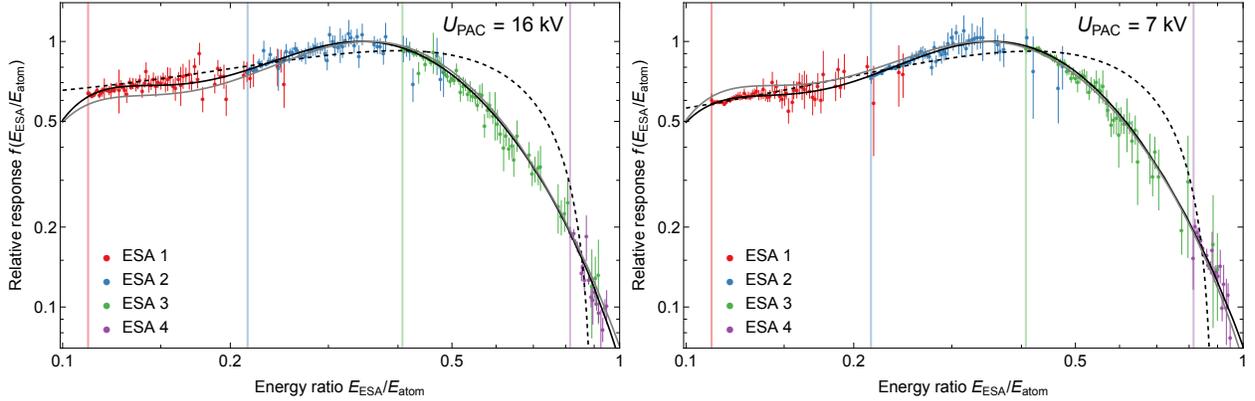

**Figure 7.** The relative response functions obtained from our analysis (solid black line) compared with the response obtained by Schwadron et al. (2022) (dashed line) and results of Equation (14) for each ESA step (points with error bars). Vertical lines indicate the lowest energy ratio for each ESA step (see text). The left and right panels show results for the PAC voltage of 16 and 7 kV, respectively. The solid gray lines show the best-fit relative response function for the other PAC voltage.

In Figure 7, we also compare our best-fit response functions with those obtained by Schwadron et al. (2022). For comparison purposes, we renormalized their functions to match our response for the energy ratio of $E_{\text{ESA}e}/E_{\text{atom}} = 0.43$. This ratio corresponds to the mean of the energy range of the data points in ESA 3 used by Schwadron et al. (2022) for their normalization to 1. The comparison shows that the response functions intersect near the clusters of points with the lowest energy ratio for each ESA step, i.e., those that correspond to the center of the energy range used in the previous analysis (see Figure 4 in Schwadron et al. 2022). However, the two functions differ between these clusters, most noticeably around the energy ratio of ~0.65. This is the energy range where the response function drops significantly with energy. However, the response function could not be constrained effectively in the previous analysis due to the very limited energy range of ISN helium observations in the ecliptic that were used in that analysis.

Moreover, the figure reveals differences between the relative response functions for the two PAC voltages, as can be seen from the additional solid gray lines added to each of the two panels, representing the response curve from the PAC value shown in the other panel. The most significant difference is evident for the lowest $E_{\text{ESA}e}/E_{\text{atom}}$ ratios, which correspond to sputtered ions with low energies relative to the incoming atom energy. These ions are likely to emerge from the conversion surface at large scattering angles. Consequently, the higher PAC voltage may result in relatively higher efficiency for their detection thanks to the focusing effect of the post acceleration. The observed differences at higher relative energy ratios (close to 1) are less statistically significant because this range includes only data points with large uncertainties.

The response function obtained from our analysis seems to feature a somewhat bimodal structure, with a clear maximum near the energy ratio of 0.35, with additional flattening near 0.13. This structure may be related to two regimes of the sputtering process: the single knock-on regime and the collision-cascade regime (Gnaser 2007). The knock-on regime is the main process allowing for observation of low energy light atoms, e.g., ISN helium atoms. In this regime, an isolated collision of the incoming atom sputters a single ion from the conversion surface. However, some probability for collision cascade may also create lower energy ions sputtered from the conversion surface. Nevertheless, further investigation of the sputtering is required to confirm this observation.

## 6. Robustness of the Relative Calibration

As discussed in Section 4, we limit the range of the data used in the analysis to minimize the potential impact from ISN hydrogen contribution for the orbits that image the proximity of the inflow direction. Nevertheless, we repeat the analysis, including those orbits, to verify the robustness of our derivation. Figure 8 shows the ratio of the obtained response function with these orbits to the results shown in Section 5. The relative change for the PAC voltage of 16 kV, which covers the initial years at solar minimum with the highest ISN H fluxes, is larger than that for the PAC at 7 kV but does not exceed 4% over the energy range covered by the data. The most substantial change occurs at the lowest energy ratios, i.e., where the ESA 1 contribution is the strongest, as is expected from additional contributions from ISN hydrogen. The change for the PAC voltage of 7 kV is smaller, up to ~1%.

We only use the model of the ISN helium population at IBEX to find the mean energy of atoms in each energy bin. The flux from the model is not used in this analysis. As discussed in Section 3, we expect that the mean energy in each bin does not strongly depend on the parameters of the ISN helium modeling. To verify that, we use the results of such modeling with modified parameters of the global heliosphere model. We considered variations in important parameters, such as the pristine VLISM flow direction and speed, temperature, interstellar magnetic field direction, and densities of plasma and neutrals. The most significant changes in the mean energy are expected for a change in the pristine VLISM speed, which is understandable because this parameter directly shifts the distribution function of ISN helium to higher speeds. We find that a change of this velocity by +0.4 km s$^{-1}$ changes the mean velocity in the spacecraft frame by at most +0.21 km s$^{-1}$, consistently with the consideration presented in Section 3. Figure 8 shows that the relative response function changes by less than ~1% compared to the nominal case, except under the PAC voltage of 16 kV for a high energy ratio, where the discrepancy increases at most to ~3%.

Finally, we verified whether changes to the ISN helium distribution function due to angular scattering in charge exchange and elastic collisions impact the results. For this purpose, we used the mean energy of ISN helium atoms calculated using the analytical full integration model (aFINM; Schwadron et al. 2013, 2015, 2016; Rahmanifard et al. 2019) calculated in Swaczyna et al. (2023). These calculations used the distribution functions estimated at the heliopause from transport in the outer heliosheath that include the momentum transfer related to charge exchange and elastic collisions. Figure 8 shows that the inclusion of this effect modifies the response function by at most ~2%

The tests discussed in this section show that the obtained result is robust. The obtained relative response function is not significantly affected by the selection of the data used for the analysis and by the assumed pristine VLISM parameters used to calculate the incoming atom energies in the spacecraft frame. Nevertheless, the assumptions introduced in Section 2 require laboratory verification and may possibly be a source of systematic error.

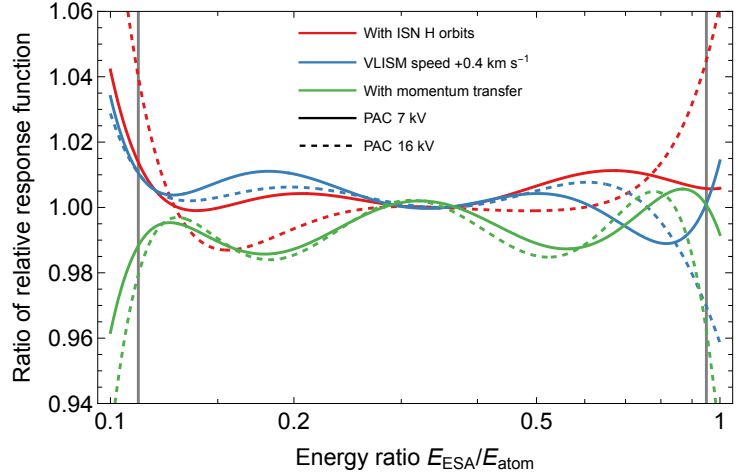

**Figure 8.** Ratios of the relative response function obtained in the robustness tests relative to the nominal case discussed in Section 5. The red lines show the results that include the orbits with possible ISN hydrogen contribution. The blue lines show the results when the assumed pristine VLISM speed was increased by +0.4 km s$^{-1}$. The green lines show the results that account for momentum transfer in collisions in the outer heliosheath. The solid and dashed lines show results obtained under the PAC voltage of 7 kV and 16 kV, respectively. The vertical lines bound the range of observed energy ratios.

## 7. Summary

IBEX-Lo detects ISN helium atoms through sputtering processes at the instrument conversion surfaces, which produce a broad spectrum of sputtered ions. Consequently, the ions are registered in all ESA steps covering energies below the incoming atom energy. Therefore, the registration in a specific ESA step does not represent the actual energy of incoming helium atoms. In our analysis, we characterize the relative response function in different ESA steps using a common function of the ratio of the nominal ESA energy to the incoming atom energy. We find the parameters of this function using ratios of observed count rates between different ESA steps as a function of the mean energy of ISN helium atoms in the IBEX frame. The mean energy was calculated using a transport model of ISN helium atoms from the pristine VLISM to 1 au accounting for their kinetic energy change due to solar gravity and for charge exchange and ionization losses in the heliosphere and the outer heliosheath based on the flows obtained from a global heliosphere model. We performed a separate analysis for each of the two PAC voltages used during the IBEX mission. We verified that the obtained response functions do not strongly depend on the parameters of this model and on the selection of the data that may include contributions from ISN hydrogen atoms.

The response function obtained from this analysis can be used to combine the observations of ISN helium from different ESA steps in quantitative analyses of IBEX-Lo observations aiming to find the pristine conditions in the VLISM. ESA steps 1 through 3 show comparable count rates, while ESA 4 has at least an order of magnitude lower rates. Nevertheless, information from ESA 4 was critical in our study to fully characterize the response function for high ratios of the ESA step energy to the incoming atom energy, i.e., where the response drops rapidly. While the IBEX-Lo team decided to eliminate ESA 4 observations during the ISN helium seasons since 2016, reintroducing this ESA step would allow for improving the knowledge of the relative response function. The knowledge of the response function is especially important for analyses that aim to find non-Maxwellian properties of the ISN helium distribution because these signatures are mostly visible in the tails, which are typically seen at higher latitudes where the incoming atom energy is lower, and thus, the ratio of the ESA step energy to the incoming atom energy is high (Sokół et al. 2015a; Swaczyna et al. 2019). Note that while we only used observations in ESA 4 for high-energy atoms, the same portion of the relative response function corresponds to low-energy atoms in ESA 3 (see Figure 7).

Our analysis shows that the relative response function in different ESA steps can be reproduced from in-flight observations. However, this relative response does not include the change in the response related to sputtering efficiency, which also depends on the incoming atom energy. Moreover, this analysis cannot be used to retrieve the absolute calibration of the instrument. Therefore, these two elements should be prioritized for future ISN detectors, e.g., as the planned IMAP-Lo instrument for Interstellar Mapping and Acceleration Probe (IMAP) mission (McComas et al. 2018). IMAP-Lo includes a moveable platform allowing for adjusting the boresight position relative to the spacecraft spin axis. This capability allows for improved observation of different ISN species (Sokół et al. 2019b) and should allow for breaking degeneracy of the VLISM parameters derived from the IBEX observations (Bzowski et al. 2022; Schwadron et al. 2022). This capability also allows for observations of high ISN helium rates over a more extensive range of incoming atom energies. Consequently, a similar analysis with IMAP-Lo observations will enable a more accurate characterization of the relative response function.

*Acknowledgments*: This material is based upon work supported by the National Aeronautics and Space Administration (NASA) under grant No. 80NSSC20K0781 issued through the Outer Heliosphere Guest Investigators Program. This work was also partially funded by the IBEX mission as a part of the NASA Explorer Program (80NSSC20K0719), and by IMAP as part of the Solar Terrestrial Probes Program (80GSFC19C0027). M.B. and M.A.K. were supported by Polish National Science Centre grant 2019/35/B/ST9/01241.